\documentclass{article}

\usepackage{PRIMEarxiv}

\usepackage[utf8]{inputenc} 
\usepackage[T1]{fontenc}    
\usepackage{hyperref}       
\usepackage{url}            
\usepackage{booktabs}       
\usepackage{amsfonts}       
\usepackage{nicefrac}       
\usepackage{microtype}      
\usepackage{lipsum}
\usepackage{fancyhdr}       
\usepackage{graphicx}       
\graphicspath{{media/}}     
\usepackage{amsmath}
\usepackage{amssymb,amsfonts}
\usepackage{multicol}

\DeclareMathOperator*{\argmin}{argmin}

\title{Nearest Subspace Search in The Signed Cumulative Distribution Transform Space for 1D Signal Classification
\thanks{~\copyright~2021 IEEE. Personal use of this material is permitted. Permission from IEEE must be obtained for all other uses, in any current or future media, including reprinting/republishing this material for advertising or promotional purposes, creating new collective works, for resale or redistribution to servers or lists, or reuse of any copyrighted component of this work in other works.} 
}

\author{
Abu Hasnat Mohammad Rubaiyat$^1$, Mohammad Shifat-E-Rabbi$^2$, Yan Zhuang$^1$, Shiying Li$^2$,\\ \textbf{Gustavo K. Rohde}$^{1,2}$\\
$^1$Department of Electrical and Computer Engineering, University of Virginia, USA\\
$^2$Department of Biomedical Engineering, University of Virginia, USA

}

\begin{document}
\maketitle

\begin{abstract}
This paper presents a new method to classify 1D signals using the signed cumulative distribution transform (SCDT). The proposed method exploits certain linearization properties of the SCDT to render the problem easier to solve in the SCDT space. The method uses the nearest subspace search technique in the SCDT domain to provide a non-iterative, effective, and simple to implement classification algorithm. Experiments show that the proposed technique outperforms the state-of-the-art neural networks using a very low number of training samples and is also robust to out-of-distribution examples on simulated data. We also demonstrate the efficacy of the proposed technique in real-world applications by applying it to an ECG classification problem. The python code implementing the proposed classifier can be found in \textit{PyTransKit}\cite{pytranskit}.
\end{abstract}


\section{Introduction}
\label{sec:intro}
Signal classification refers to the automatic prediction of the class label of an unknown signal using the information extracted from the corresponding signal values. It is at the heart of many signal processing applications, e.g. human activity recognition \cite{lara2012survey}, physiological signal classification \cite{berkaya2018survey}\cite{subasi2010eeg}, communication \cite{fehske2005new}, machine health monitoring systems \cite{zhao2019deep}, financial time series data \cite{lu2009financial}, etc. In the biomedical domain, the classification of physiological signals, such as ECG, EEG, plays an important role in the diagnosis of several diseases.

Existing signal classification approaches can be categorized into two broad groups: 1) hand-engineered feature-based classifiers and 2) end-to-end learning-based classifiers such as artificial neural networks. Feature based methods \cite{phinyomark2012feature}\cite{luz2016ecg} usually employ a two-step process: first, extraction of numerical features, e.g. time or frequency domain features, wavelet features, etc. from the raw signal data, and then,  application of different multivariate regression-based classification methods, e.g. linear discriminant analysis, support vector machines, random forests, etc. on the extracted features. Convolutional neural networks (CNN) based classification methods have widely been studied recently as they have shown promising results in certain classification tasks \cite{hershey2017cnn} \cite{yildirim2018arrhythmia}.  

A less commonly used alternative is to use transport transform based classification techniques \cite{kolouri2017optimal}. Following the linear optimal transport framework proposed by Wang et al. \cite{wang2013linear}, the cumulative distribution transform (CDT)  \cite{Park:18} was proposed as a means of classifying strictly positive 1D signals. Aldroubi et al. \cite{aldroubi2021signed} proposed the signed cumulative distribution transform (SCDT), an extension of the CDT to general signed signals. Both transforms can be applied to 1D signal classification problems in combination with linear classifiers, such as Fisher discriminant analysis, support vector machines, etc. In this paper, we propose a new classification technique for 1D signals that first represents each signal in the SCDT space, and following \cite{shifat2020radon} we then utilize the nearest subspace method to classify each signal. Using a well-known ECG signal database, we show the approach is efficient to compute, highly accurate, requires few training samples, and performs well in out-of-distribution tasks as compared to state of the art end-to-end convolutional neural networks.

\section{Proposed Method}
\label{sec:method}
\vspace{-0.5em}
\subsection{The Signed Cumulative Distribution Transform}
\label{ssec:SCDT}
The signed cumulative distribution transform (SCDT) \cite{aldroubi2021signed} is an extension of the cumulative distribution transform (CDT) \cite{Park:18}, an invertible nonlinear 1D signal transform from the space of smooth positive probability densities to the space of diffeomorphisms. The CDT was introduced for a class of normalized positive smooth functions, which can be described as follows: let $s(t), t\in\Omega_s$ and $s_0(y),y\in\Omega_{s_0}$ define a given signal and a (known) reference signal, respectively, such that $\int_{\Omega_s}s(u)du = \int_{\Omega_{s_0}}s_0(u)du = 1$ and $s_0(y), s(t)>0$ in their respective domains. The CDT of $s(t)$ is then defined to be the function $s^*(y)$ that solves,
\begin{align}
    \int_{\inf(\Omega_s)}^{s^*(y)} s(u)du = \int_{\inf(\Omega_{s_0})}^{y} s_0(u)du.
    \label{eq:cdt}
\end{align}
Now if we define the cumulative distribution functions (CDFs) $S(t) = \int_{-\infty}^{t} s(u)du$ and $S_0(y) = \int_{-\infty}^{y} s_0(u)du$, an alternative expression for $s^*(y)$ is given by,
\begin{equation}
	s^*(y) = S^{-1}(S_0(y)).
	\label{eq:cdt_alt}
\end{equation}
If a uniform reference signal is used, i.e. $s_0(y) = 1$ in $\Omega_{s_0}=[0,1]$, we can write $S_0(y) = y$ and therefore $s^*(y) = S^{-1}(y)$. Note that we are using the definition of the CDT used in \cite{Rubaiyat:20}, which is slightly different from the formulation used in \cite{Park:18}.

Although the CDT can effectively be used in many classification \cite{Park:18} and estimation \cite{Rubaiyat:20} problems, the framework described above has the limitation that signals themselves must be positive. To overcome this limitation, Aldroubi et al. \cite{aldroubi2021signed} proposed the SCDT as an extension of the CDT to general finite signed signals. For the non-negative signal $s(t)$ with arbitrary mass, the transform is given by:
\begin{equation}
    \widehat{s}(y) = \begin{cases}
    \left(s^*(y),\|s\|_{L_1}\right),& \text{if } s\neq 0\\
    (0,0),              & \text{if } s=0,
\end{cases}
\label{eq:scdt_mass}
\end{equation}
where $\|s\|_{L_1}$ is the $L_1$ norm of signal $s$, and $s^*$ is simply the CDT (eq. (\ref{eq:cdt_alt})) of the normalized signal $\frac{1}{\|s\|_{L_1}}s$. 

Now to extend the transform to a signed signal, the signal $s(t)$ is decomposed as $s(t) = s^+(t) - s^-(t)$, where $s^+(t)$ and $s^-(t)$ are the absolute values of the positive and negative parts of the signal $s(t)$. The SCDT of $s(t)$ is then defined as:
\begin{equation}
    \widehat{s}(y) = \left(\widehat{s}^+(y), \widehat{s}^-(y)\right),
    \label{eq:scdt}
\end{equation}
where $\widehat{s}^+(y)$ and $\widehat{s}^-(y)$ are the transforms, as defined in eq. (\ref{eq:scdt_mass}), for the positive signals $s^+(t)$ and $s^-(t)$, respectively. 
The SCDT has a number of properties which will help us simplify the signal classification problems.

\textbf{Composition property:}
Let $\widehat{s}$ be the SCDT of the signed signal $s$. The SCDT of the signal $s_g=g's\circ g$ is given by:
\begin{equation}
    \widehat{s}_g = \left(g^{-1}\circ (s^+)^*,\|s^{+}\|_{L_1},g^{-1}\circ (s^-)^*,\|s^{-}\|_{L_1}\right),
    \label{eq:scdt_composition}
\end{equation}
where, $g(t)$ is an invertible and differentiable increasing function, $s\circ g = s(g(t))$, and $g'(t)=dg(t)/dt$ \cite{aldroubi2021signed}. For example, in case of translation and scaling, $g(t) = \omega t - \tau$, and the SCDT of the signal $s_g(t)=\omega s(t - \mu)$ can be derived as: $\widehat{s}_g = \left(\frac{(s^+)^* + \mu}{\omega},\|s^{+}\|_{L_1},\frac{(s^-)^* + \mu}{\omega},\|s^{-}\|_{L_1}\right)$.


\textbf{Convexity property:}
Let a set of signals be defined as: $\mathbb{S}=\{s_j|s_j=g'_j \varphi\circ g_j, g_j\in \mathcal{G}\}$, where $\varphi$ is a given signal and $\mathcal{G}\subset \mathcal{T}$ denotes a set of 1D spatial deformations of a specific kind (e.g. translation, scaling, etc.). Convexity property states that the set $\widehat{\mathbb{S}} = \{\widehat{s}_j:s_j\in \mathbb{S}\}$ is convex for every $\varphi$ if and only if $\mathcal{G}^{-1}=\{g_j^{-1}:g_j\in \mathcal{G}\}$ is convex \cite{aldroubi2021signed}. Note that $\mathcal{T}$ is a set of all possible increasing diffeomorphisms from $\mathbb{R}$ to $\mathbb{R}$.

The set $\mathbb{S}$ can be interpreted as a generative model for a signal class while $\varphi$ being the template signal for that class. In the next section, we discuss the generative model based problem formulation for 1D signal classification problem. Later, we show how the composition and convexity properties of the SCDT help render certain nonlinear signal classes into convex clusters and facilitate simple classification algorithms.
\vspace{-1.0em}

\subsection{Generative Model and Problem Statement}
\label{ssec:genModel}
This section describes a generative model-based problem statement for the 1D signal classification problems. We are concerned with classifying signal patterns that can be modeled as the instances of a certain template observed under some unknown deformations (e.g., translation, dispersion, etc.). For example, we can consider classifying ECG signals for detecting abnormal heart conditions. The ECG segments of a class can be thought of as instances of a template corresponding to that class, each observed under some deformations. We define the signal classes of the above type with the following generative model:

\textbf{Generative model:} Let $\mathcal{G}\subset \mathcal{T}$ denotes a set of increasing 1D spatial deformations of a specific kind (e.g., $g(t)=\sum_{k}p_kt^k \in \mathcal{G}$, $g'(t)>0$). The 1D mass (signal intensity) preserving generative model for class-$c$ is then defined to be the set:
\begin{align}
    \mathbb{S}^{(c)}=\{s_j^{(c)}|s_j^{(c)}=g'_j\varphi^{(c)}\circ g_j, g_j\in \mathcal{G}\},
    \label{eq:gen_mod}
\end{align}
where $\varphi^{(c)}$ is the template pattern of class $c$, $s_j^{(c)}$ is the $j$-th signal from that class, and $g'_j>0$. Considering the generative model, the classification problem is as follows:

\textbf{Classification problem:} \textit{Let $\mathcal{G}\subset \mathcal{T}$ be the set of spatial deformations and $\mathbb{S}^{(c)}$ be defined as in eq. (\ref{eq:gen_mod}). Given training samples $\{s_1^{(c)}, s_2^{(c)}, ...\}$ for class-$c$ ($c=0, 1, 2, ..., N$), determine the class label of an unknown signal $s$}.

\subsection{Proposed Solution}
\label{ssec:solution}
The generative model described in eq. (\ref{eq:gen_mod}) generally yields nonconvex (hence nonlinear) signal classes. However, the SCDT can simplify the data geometry and thereby simplify the classification problem. Therefore we follow the approach proposed in \cite{shifat2020radon} and employ the composition property of the SCDT on $s_j^{(c)}$ in eq. (\ref{eq:gen_mod}) such that
\begin{align*}
    \widehat{s}_j^{(c)} = \left(g_j^{-1}\circ (\varphi^{(c)^+})^*,\|\varphi^{(c)^+}\|_{L_1},g_j^{-1}\circ (\varphi^{(c)^-})^*,\|\varphi^{(c)^-}\|_{L_1}\right),
\end{align*}
where, $(\varphi^{(c)^+})^*$ and $(\varphi^{(c)^-})^*$ are the CDTs (defined in eq. (\ref{eq:cdt_alt})) of the positive and negative parts of the template signal $\varphi^{(c)}$, respectively. The constant terms $\|\varphi^{(c)^+}\|_{L_1}$ and $\|\varphi^{(c)^-}\|_{L_1}$ denote the corresponding total masses. Noting that signals can be made to have zero mean, in the proposed solution we ignore the total mass terms (constant scaling factor) and focus on the CDT terms only, which yields a modified transformed signal:
\begin{align}
    \widehat{s}_j^{(c)\dagger} = \left(g_j^{-1}\circ (\varphi^{(c)^+})^*, g_j^{-1}\circ (\varphi^{(c)^-})^*\right).
    \label{eq:scdt_g}
\end{align}
The generative model in the transform domain is given by,
\begin{equation}
\begin{split}
    \widehat{\mathbb{S}}^{(c)} = \{\widehat{s}_j^{(c)\dagger}| \widehat{s}_j^{(c)\dagger}=\left(g_j^{-1}\circ (\varphi^{(c)^+})^*, g_j^{-1}\circ (\varphi^{(c)^-})^*\right),\\
    g_j\in \mathcal{G}\}.
    \label{eq:gen_mod_scdt}
\end{split}
\end{equation}
It can be shown that the generative model given in eq.~\eqref{eq:gen_mod_scdt} forms a convex set if $\mathcal{G}$ is a convex group \cite{shifat2020radon}. 
Moreover, as the SCDT is a one-to-one map, it follows that if $\mathbb{S}^{(c)} \cap \mathbb{S}^{(p)}=\varnothing$, then $\widehat{\mathbb{S}}^{(c)} \cap  \widehat{\mathbb{S}}^{(p)}=\varnothing$. Now, let us define a subspace generated by the convex set $\widehat{\mathbb{S}}^{(c)}$ as follows: 
\begin{equation}
    \widehat{\mathbb{V}}^{(c)} = \text{span}\left(\widehat{\mathbb{S}}^{(c)}\right) = \left\{\sum_{j\in \mathbf{J}}\alpha_j\widehat{s}_j^{(c)\dagger}|\alpha_j\in\mathbb{R},\mathbf{J} \text{ is finite} \right\}.
    \label{eq:subspace}
\end{equation}
It can be shown that under certain assumptions, the convex space for a particular class in transform domain does not overlap with the subspace corresponding to a different class \cite{shifat2020radon}, i.e. $\widehat{\mathbb{S}}^{(c)}\cap\widehat{\mathbb{V}}^{(p)}=\varnothing,c\neq p$. 

It follows from the analysis that for a test sample $s$ generated according to the generative model for class $c$ (unknown), $d^2( \widehat{s}^\dagger,\widehat{\mathbb{V}}^{(c)}) \sim 0$, and $d^2( \widehat{s}^\dagger,\widehat{\mathbb{V}}^{(p)})>0$ for $p\neq c$. Here $d^2(\cdot,\cdot)$ is the Euclidean distance between $\widehat{s}$ and the nearest point in $\widehat{\mathbb{S}}^{(c)}$ or $\widehat{\mathbb{V}}^{(c)}$, and $\widehat{s}^\dagger = \left((s^{+})^*, (s^{-})^* \right)$. Therefore, the class of the unknown sample can be predicted by solving,
\begin{equation}
    \argmin_c~d^2(\widehat{s}^\dagger, \widehat{\mathbb{V}}^{(c)}). \nonumber
    \label{eq:soln1}
\end{equation}

\begin{figure}[!tb]
\centering
\centerline{\includegraphics[width=0.55\textwidth]{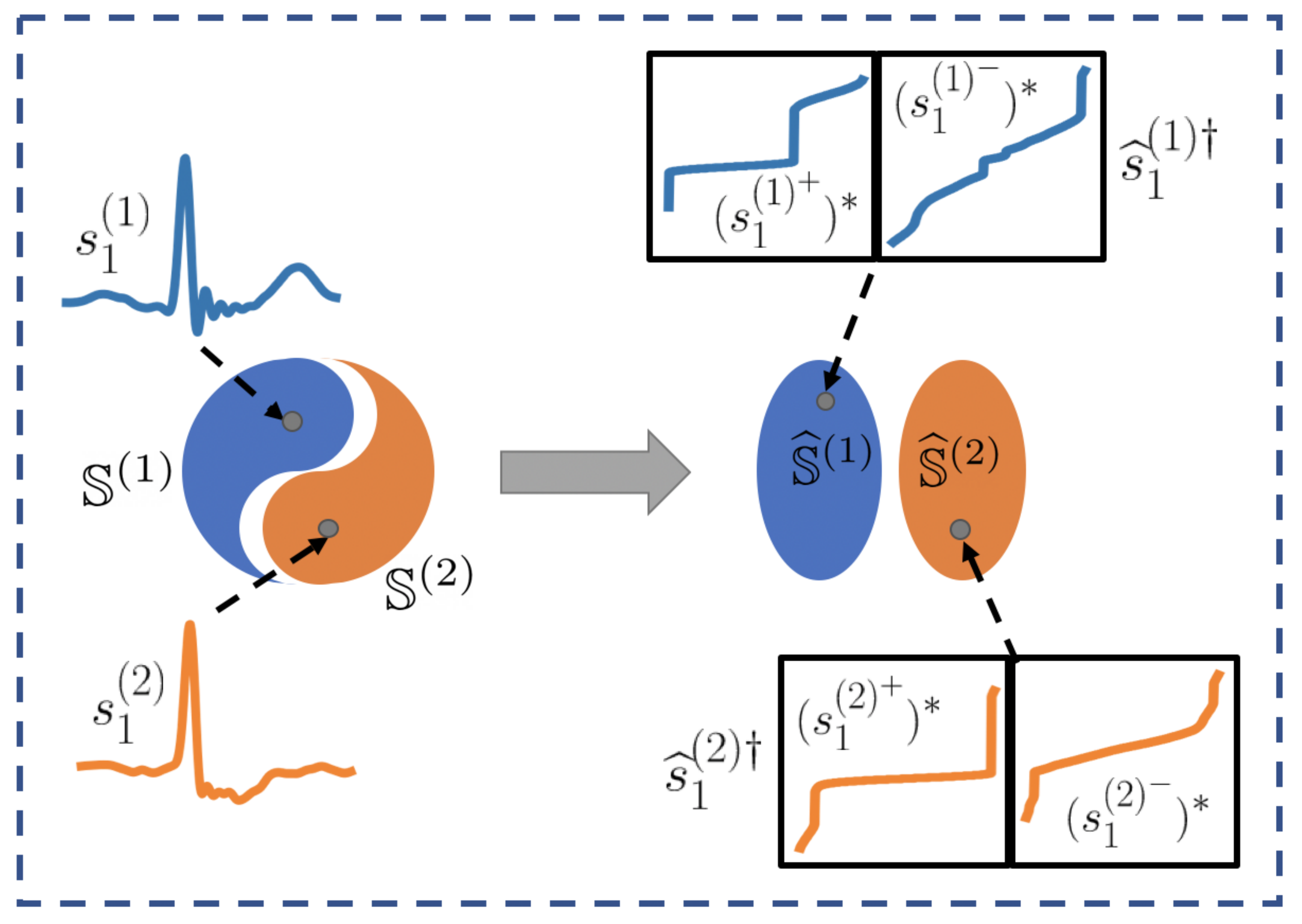}}
\caption{First step of the training algorithm is to obtain the transform space representations $\widehat{\mathbb{S}}^{(c)}$ of the given training samples for class $c$. Later, $\widehat{\mathbb{S}}^{(c)}$ is orthogonalized to obtain the basis vectors that span the space $\widehat{\mathbb{V}}^{(c)}$.}
\label{fig:train_alg}
\end{figure}

\textbf{Training Algorithm:}
Based on the analysis discussed above, we can follow \cite{shifat2020radon} to propose a non-iterative training algorithm.
First, we compute the transforms of the training samples to get $\widehat{\mathbb{S}}^{(c)}$ for all the classes (Fig. \ref{fig:train_alg}). Next, we approximate $\widehat{\mathbb{V}}^{(c)} = \text{span}\left(\widehat{\mathbb{S}}^{(c)} \right)=\text{span}\left\{\widehat{s}_1^{(c)\dagger}, \widehat{s}_2^{(c)\dagger}, ... \right\}$. We orthogonalize $\{\widehat{s}_1^{(c)\dagger}, \widehat{s}_2^{(c)\dagger}, ... \}$ to obtain the set of basis vectors $\{b_1^{(c)}, b_2^{(c)}, ... \}$ that spans the space $\widehat{\mathbb{V}}^{(c)}$ for class $c$. We then form the matrix $B^{(c)}$ with the computed basis vectors in its columns: $B^{(c)} = \left[b_1^{(c)}, b_2^{(c)}, ...  \right]$, which will later be used to predict the class label of an unknown sample. 
Note that although this algorithm is designed for classifying signal patterns that can be modeled  as  the  instances  of  a  certain  template (see the generative model defined in eq. \ref{eq:gen_mod}), we do not need to estimate the template $\varphi^{(c)}$ for class $c$. One can simply take the training samples for a particular class $c$ and generate $B^{(c)}$ using the algorithm described above.

\textbf{Testing Algorithm:}
Let us consider the problem of predicting class of a test signal $s$ using the proposed model. First, we obtain the transform $\widehat{s}^\dagger = \left((s^{+})^*, (s^{-})^* \right)$. We then estimate the distance between $\widehat{s}^\dagger$ and the subspace model for each class by $d^{2}(\widehat{s}, \widehat{\mathbb{V}}^{(c)})\sim \| \widehat{s} - B^{(c)}{B^{(c)}}^T\widehat{s} \|_{L_2}^2$. The class of $\widehat{s}^\dagger$ is estimated to be: 
\begin{equation}
    \argmin_c~\|\widehat{s}^\dagger - A^{(c)}\widehat{s}^\dagger\|_{L_2}^2,
    \label{eq:test_alg}
\end{equation} 
where $A^{(c)} = B^{(c)}{B^{(c)}}^T$ is an orthogonal projection matrix onto the subspace spanned by the columns of $B^{(c)}$. 

\section{Experiments and Results}
\label{sec:exp} \vspace{-0.5em}
\subsection{Experimental Setup}
\label{ssec:setup}
In this section, we evaluate the performance of the proposed method with respect to some popular CNN-based 1D signal classification techniques. To demonstrate the ability of the proposed classifier, we first consider the problem of classifying three prototype signal classes: a Gabor wave, an apodized sawtooth wave, and an apodized square wave. We then conducted experiments on the synthetic data and compared the results against three different CNNs: a deep \cite{schirrmeister2017deep}, a shallow \cite{schirrmeister2017deep}, and a compact \cite{lawhern2018eegnet} CNN. Classification performances of the methods were studied in terms of test accuracy, data efficiency, and robustness to the out-of-distribution samples. We also used an ECG dataset to demonstrate the comparative performances of the methods on real data.
\vspace{-1.0em}

\subsection{Evaluation}
\label{ssec:eval} \vspace{-0.5em}
\subsubsection{Effective and Data Efficient}
\begin{figure}[!tb]
\centering
\centerline{\includegraphics[width=0.65\textwidth]{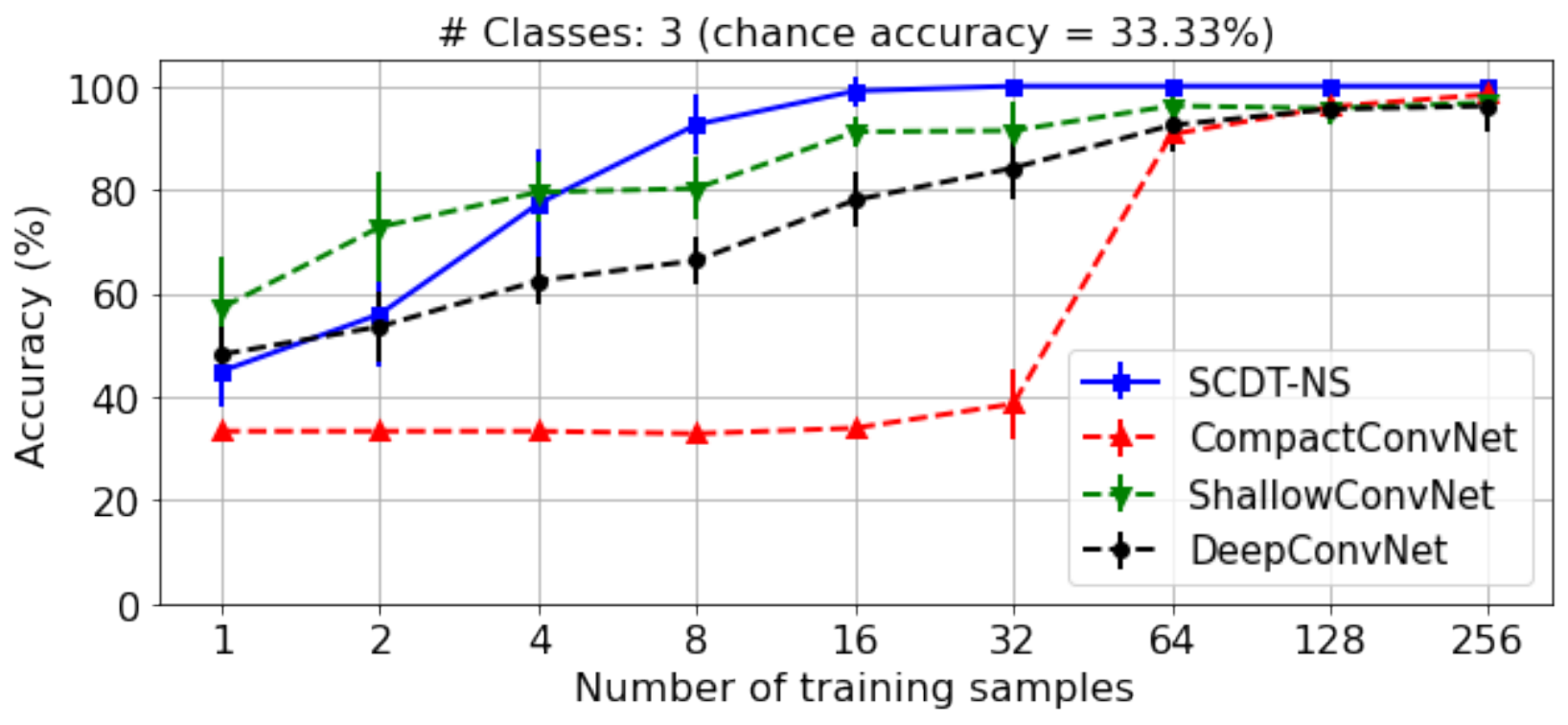}}\vspace{-1.0em}
\caption{Test accuracy on the synthetic dataset as a function of the number of training samples per class.}
\label{fig:tp_res}
\end{figure}

\vspace{-0.5em}
To demonstrate the effectiveness and data efficiency of the proposed method, we generated a synthetic dataset by applying $4$th degree polynomials on the three aforementioned prototype signals. Test results plotted in Fig. \ref{fig:tp_res} show that the proposed method required very few train samples to learn the deformations present in the data, as it achieved near perfect accuracy with only $16$ train samples per class. From the figure, it is also evident that none of the CNNs attained perfect accuracy even with a training set as large as $256$ samples per class. This experiment shows that the proposed method requires very few training data to outperform the deep learning based classification methods.

\subsubsection{Robust to Out-of-Distribution Samples}
\begin{figure}[!tb]
\centering
\centerline{\includegraphics[width=0.65\textwidth]{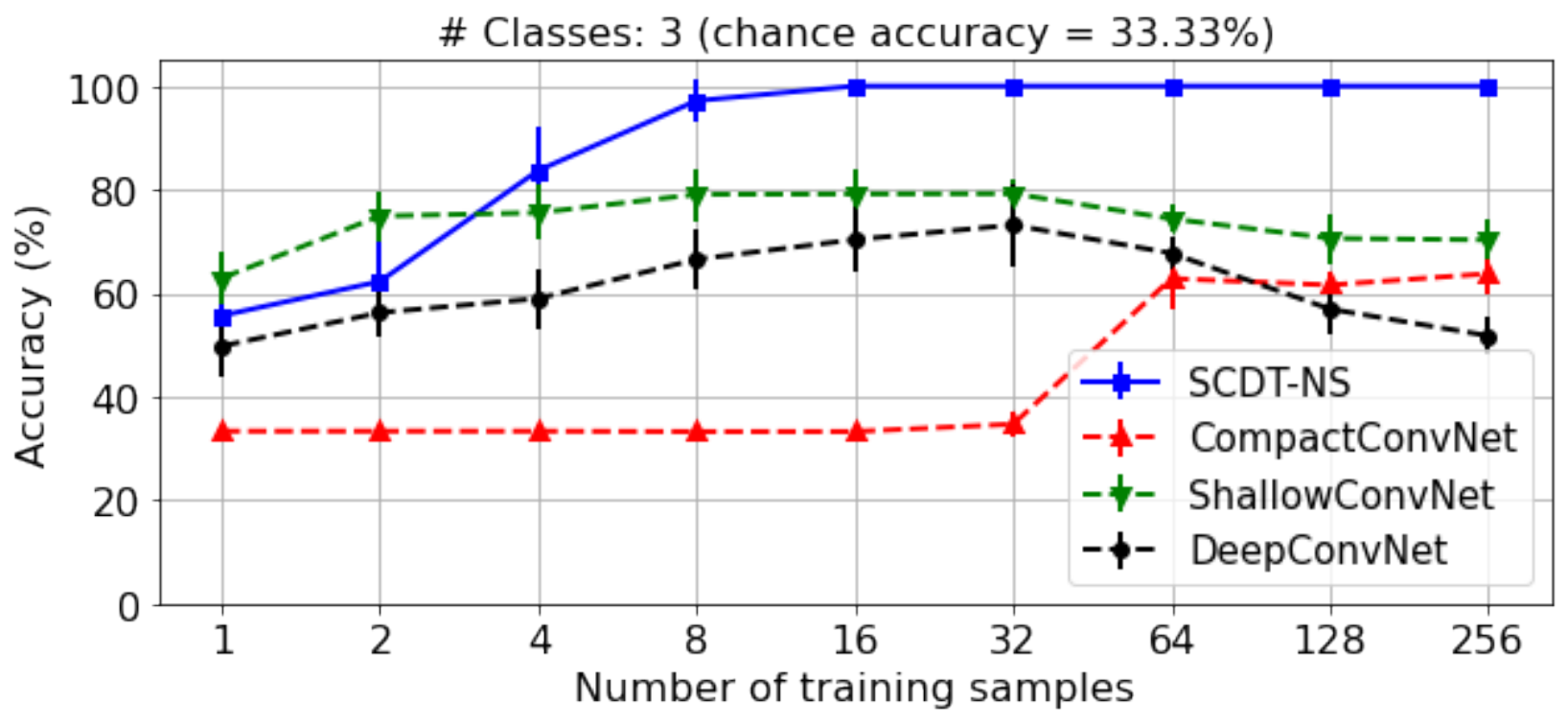}}\vspace{-1.0em}
\caption{Test accuracy as a function of the number of training samples per class under the out-of-distribution setup. The out-of-distribution setup consists of ‘in distribution’ training and ‘out distribution’ testing sets containing different sets of magnitudes of the confounding factors.}
\label{fig:tp_outdist}
\end{figure}

To demonstrate the robustness to out-of-distribution examples, we used $4$th degree polynomials to generate data, but varied the magnitude of the confounding factors (i.e. the coefficients of the polynomials) to generate a gap between training and testing distributions. The `in distribution' set used during the training process consisted of signals with relatively smaller confounding factors (smaller coefficient values) than the `out distribution' set used for testing. A similar concept was used in \cite{shifat2020radon} to perform out-of-distribution testing. Fig. \ref{fig:tp_outdist} shows that the proposed method outperformed the other methods by a greater margin than before under this modified experimental scenario.

\subsection{Application: ECG Signal Classification}
\label{ssec:ecgData}
The proposed SCDT-NS classification technique can be applied to classify 1D signals in many applications. One such application is electrocardiogram (ECG) signal classification. To evaluate the performance of the proposed model on ECG data, we used a publicly available dataset \cite{ecg744} reported in \cite{plawiak2018novel}. This dataset was collected from MIT-BIH Arrhythmia database \cite{moody2001impact} hosted at PhysioNet \cite{goldberger2000physiobank}. Although, the dataset has $17$ different classes, in this paper we used $3$ classes with three highest number of ECG fragments: 1) Normal sinus rhythm (NSR), 2) Atrial fibrillation (AFIB), and 3) Left bundle branch block beat (LBBBB). PQRST locations were detected using the method described in \cite{bassiouni2018intelligent} to segment the heartbeats from the ECG fragments. Then we applied the classification methods to classify the ECG heartbeat segments. Note that during experiments, we made sure data from same patients was not included in both training and test sets. This step is often not strictly followed in the computer science (machine learning) literature. Mixing data from the same patient in both the test and training samples tends to lead to high accuracies ($\sim 97\%$), which is attained by all the methods compared here. Table \ref{table:ecg_res} summarizes the comparative results among the classification methods. It shows that the proposed SCDT-NS method outperformed the CNN methods in classifying ECG heartbeat signals.

\begin{table}[h!]
\centering
\begin{tabular}{ | l | c | c | } 
  \hline
  & \textbf{Accuracy (\%)} & \textbf{F1 score} \\ 
  \hline
  DeepConvNet &  47.57& 0.4065 \\ 
  \hline
  ShallowConvNet &  33.68& 0.2618 \\ 
  \hline
  CompactConvNet &  29.59& 0.2466 \\ 
  \hline
  \textbf{SCDT-NS} &  \textbf{61.50}& \textbf{0.5979} \\ 
  \hline
\end{tabular}
\caption{Accuracy and F1 scores for different classification methods on ECG data. (trained with $512$ ECG heartbeat segments per class, tested on $300$ samples per class.)}
\label{table:ecg_res}
\vspace{-0.5em}
\end{table}

\section{Conclusion}
\label{sec:conc}
\vspace{-1.0em}
In this paper, we introduced a new method for 1D signal classification problems. The proposed approach employs a nearest subspace search algorithm in the signed cumulative distribution transform (SCDT) space to produce a non-iterative solution to the classification problem. Several experiments show that the proposed method not only outperforms popular CNN-based methods but also is data efficient and robust to out-of-distribution samples. Moreover, it does not require prior knowledge of the templates of the signal classes or the deformations present in the data to render better accuracy than the comparative methods. Future work will include studying ways to specify general mathematical categories for the space of signal deformations $\mathcal{G}$, as well as way to learn them more efficiently.

\section*{Acknowledgments}
This work was supported in part by NIH grants GM130825, GM090033.

\bibliographystyle{unsrt}  
\bibliography{main}

\end{document}